\documentclass[aps,superscriptaddress,twocolumn]{revtex4}%
\usepackage{amsfonts}%
\usepackage{amsmath}%
\usepackage{amssymb}%
\usepackage{graphicx}
\usepackage{color}

\newtheorem{theorem}{Theorem}

\newtheorem{corollary}[theorem]{Corollary}

\newtheorem{lemma}[theorem]{Lemma}

\newenvironment{proof}[1][Proof]{\noindent\textbf{#1.} }{\ \rule{0.5em}{0.5em}}
\begin{document}

\title{Random subspaces for encryption based on a private shared Cartesian
frame}
\author{Stephen D. Bartlett}
\affiliation{School of Physics, The University of Sydney, New South Wales 2006,
Australia}
\author{Patrick Hayden}
\affiliation{School of Computer Science, McGill University,
Montreal, Canada}
\author{Robert W. Spekkens}
\affiliation{Perimeter Institute for Theoretical Physics, Waterloo,
Ontario N2L 2Y5, Canada}
\date{23 November 2005}

\begin{abstract}
  A private shared Cartesian frame is a novel form of private shared
  correlation that allows for both private classical and quantum
  communication.  Cryptography using a private shared Cartesian
  frame has the remarkable property that asymptotically, if perfect privacy is
  demanded, the private classical capacity is \emph{three times} the
  private quantum capacity.  We demonstrate that if
  the requirement for perfect privacy is relaxed, then it is
  possible to use the properties of random subspaces to nearly
  triple the private quantum capacity, almost closing the gap
  between the private classical and quantum capacities.
\end{abstract}

\pacs{03.67.Dd,03.67.Hk,03.65.Pp}%
\keywords{random subspaces,superselection rules, reference frames,
encryption}

\maketitle

\section{Introduction}

Quantum information theory is concerned with implementing various
communications tasks with a minimal use of resources~\cite{Nie00}.
In multi-party protocols, the most interesting resources are
\emph{nonlocal} ones, such as shared classical key or entanglement.
Recently, it has become apparent that shared reference frames (SRFs)
are another form of nonlocal resource that may be included in the
accounting of any multi-party information-processing protocol.
Heuristically, two parties are said to share a reference frame if
there is a perfect correlation between the systems that define the
bases of their respective local Hilbert spaces.  For instance, if
Alice and Bob each define their local Cartesian frame using
classical gyroscopes in their labs, then they possess a shared
reference frame if the rotation relating the frames defined by their
gyroscopes is known to them.

Like entanglement, no amount of discussion between Alice and Bob
will allow them to establish a shared reference frame; doing so
requires a physical interaction between them that goes beyond the
framework of classical information theory and, for that matter, the
usual formalism of quantum information theory.  For example, to
establish a shared Cartesian frame between their respective labs,
Alice and Bob may make use of a \emph{pre-existing} frame such as
the fixed stars or the Earth's magnetic field.  However, if no such
shared frame exists \emph{a priori}, then no amount of discussion
will enable them to establish one; to do so, they must exchanges
physical systems that carry some directional information such as
spin-$1/2$ particles.  Understanding the value of a shared reference
frame as a new nonlocal resource and its relation to both private
and quantum communication is therefore an important necessary step
in the ongoing effort to understand the nature of information in
physics.

Substantial progress has recently been made in this direction.
Most research has focussed on determining the communication cost
to establishing an
SRF~\cite{GisPop,Per01,Bag01,Per01b,Bag01b,Lin03,Chi04}.  There
have also been several investigations into the impact of SRFs (or
the lack thereof) on the efficiency with which one can perform a
variety of bi-partite tasks, such as quantum and classical
communication~\cite{BRS03a,Enk04}, key
distribution~\cite{Wal03,Boi03}, and the manipulation of
entanglement~\cite{Ver03,Bar03,BDSW04,Enk04}. Other work has
studied the cryptographic consequences of the participants' lack
of an SRF~\cite{KMP03,Ver03,HOT05}. Here we shall be interested
instead in using \emph{private} SRFs as a resource.

An SRF is private if the systems that define Alice and Bob's local
Hilbert space bases are not correlated with any other systems. In
this case, the SRF can act as a new kind of key for private quantum
and classical communication over a public channel. Consider a
private shared Cartesian frame, as in~\cite{BRS04}. Under the
requirement that the communication have perfect fidelity and the
privacy be perfect, it was found that for $N$ transmitted qubits,
the number of private qubits that can be communicated asymptotically
is $\log_2 N$, and the number of private classical bits that can be
communicated asymptotically is $3\log_2 N$.  This unusual factor of
three relating the quantum and classical capacities is understood in
terms of the details of the representation theory of
SU(2)~\cite{Ful91}, but should be contrasted with the ``usual''
factor of two that typically relates classical and quantum
schemes~\cite{Ben92,Amb00}. In the present paper, we ask the
question of whether these capacities may be improved by allowing
transmission with near-perfect rather than perfect privacy, as is
usually considered in cryptography.

Note the following suggestive facts. $2N$ secret shared classical
bits can be used in a one-time pad to encrypt $2N$ classical bits
(cbits).  However, if one asks how many private \emph{qubits} can be
transmitted using the secret key, the answer is a factor of two
less; that same secret $2N$ cbit string can only encrypt $N$ qubits
if perfect privacy is required~\cite{Amb00}. If only near-perfect
privacy is required, on the other hand, the number of secret cbits
required per encrypted qubit shrinks from $2$ to $1$
asymptotically~\cite{HLSW04}, so that the difference between
encrypting cbits and qubits disappears.

A similar effect occurs in the domain of communication using shared
entanglement. The superdense coding protocol~\cite{Ben92} uses the
transmission of $N$ qubits and the consumption of $N$ ebits to
communicate $2N$ cbits with perfect privacy.  (An ebit is a pure,
maximally entangled state of two qubits.)  The number of qubits that
can be transmitted with perfect privacy and no errors is again a
factor of two less, as implemented in the quantum Vernam
cipher~\cite{Leu02}. If either near-perfect privacy or near-perfect
transmission is permitted, however, the superdense coding protocol
can be extended to allow the transmission of nearly $2N$
\emph{qubits}~\cite{HHL04,HLW04}, again erasing the difference
between sending classical and quantum data. The fact that the
methods developed in \cite{BRS04} for communicating private
classical data using a private shared reference frame made heavy use
of superdense coding suggests that it may be possible, by relaxing
the security conditions, to increase the private quantum capacity by
a factor of three, from $\log_2 N$ to nearly $3\log_2 N$.  We shall
show that this is indeed the case.

\section{Preliminaries}

\subsection{Brief comment on notation}

The symbol for a state (such as $\varphi$ or $\rho$) also denotes
its density matrix.  A \emph{pure} state is always denoted as a ket
(e.g., $|\varphi\rangle$) and the density matrix for a pure state
$|\varphi\rangle$ is written simply as $\varphi$. We will also use
the notation $x_N \sim y_N$ if $\lim_{N\rightarrow\infty} x_N/y_N =
1$.  The term \emph{irrep} denotes an irreducible representation of
a group.

\subsection{Private quantum channels using private shared correlations}

Whenever Alice and Bob have some private shared correlation, that
is, one to which an eavesdropper Eve does not have access, Eve's
description of the systems transmitted along the channel is
related to Alice's description by a decohering superoperator,
denoted by $\mathcal{E}$~\cite{Amb00,BRS04}. Before discussing
shared reference frames specifically, we begin by formalizing the
notions of the private quantum and classical capacities of this
decohering superoperator.

A $\delta$-\emph{private quantum communication scheme for}
$\mathcal{E}$ consists of a completely positive, trace-preserving
encoding $\mathcal{C}$, mapping message states on a logical Hilbert
space $\mathbb{H}_{L}$ to encoded states on the Hilbert space
$\mathbb{H}$ of the transmitted system, such that (i) the operation
$\mathcal{C}$ is invertible by Bob (who possesses the private shared
correlations), allowing him to decode and recover states on
$\mathbb{H}_{L}$ with perfect fidelity, and (ii) the encoding
satisfies
\begin{equation}
  \left\Vert \mathcal{E}(\varphi)-\rho_{0}\right\Vert _{1}
  \leq\delta\,,\quad\forall\ |\varphi\rangle \in \mathbb{H}_{L}
  \,,
\end{equation}
where $\rho_{0}$ is some fixed state on $\mathbb{H}$, $\left\Vert
\rho -\sigma\right\Vert _{1}\equiv {\rm Tr}|\rho-\sigma|$ is the
trace distance between $\rho$ and $\sigma$, and $\delta$ is a
security parameter. When $\delta=0$, the scheme is said to be
\emph{perfectly private}. The \emph{private quantum capacity} of
this channel, $Q(\mathcal{E},\delta)$, is defined as
$Q(\mathcal{E},\delta)= \sup_{\mathcal{C}} \log_2 {\rm dim}\,
\mathbb{H}_L$.

A $\delta$-\emph{private classical communication scheme for
}$\mathcal{E}$ consists of a set $\{\rho_{i}\}_{i=1}^m$ of density
operators on $\mathbb{H}$ such that (i) the $\{\rho_{i}\}$ are
orthogonal, so that Bob can distinguish these classical messages
with certainty, and (ii) the encoding satisfies
\begin{equation}
  \left\Vert \mathcal{E}[\rho_{i}]-\rho_{0}\right\Vert _{1}\,\leq\delta
  \quad\forall\ i\,,
\end{equation}
where, again, $\rho_{0}$ is some fixed state in $\mathbb{H}$ and
$\delta$ is a security parameter. The \emph{private classical
capacity} of this channel, $C(\mathcal{E},\delta)$, is defined to be
$C(\mathcal{E},\delta)= \sup_{\{\rho_i\}} \log_2 |\{\rho_i\}|$,
where the supremum is over sets of density operators $\{\rho_i\}$
achieving $\delta$-privacy.

Given $\delta$-privacy, for any pair of quantum or classical
messages, chosen with equal prior probabilities, the probability
that Eve can distinguish these is bounded above by $(1+\delta)/2$,
seen as follows. Suppose the message states are $\varrho_{1}$ and
$\varrho_{2}$. These could be either an encoded pair of quantum
messages, that is, $\mathcal{C}(\varrho_{L,1})$ and
$\mathcal{C}(\varrho_{L,2})$ for some pair of density operators
$\varrho_{L,1}$ and $\varrho_{L,2}$ on $\mathbb{H}_L$, or an
encoded pair of classical messages, that is, orthogonal density
operators. In either case, the optimal probability for Eve to
distinguish $\mathcal{E}(\varrho_{1})$ and
$\mathcal{E}(\varrho_{2})$ is given by
$\frac{1}{2}+\frac{1}{4}\left\Vert
\mathcal{E}(\varrho_{1})-\mathcal{E}(\varrho_{2})\right\Vert
_{1}$~\cite{Hel76,Fuc98}. Making use of the triangle inequality
for the trace norm $\|\cdot\|_1$, we obtain
\begin{align}
  \Vert \mathcal{E}(\varrho_{1})&-\mathcal{E}(\varrho_{2})\Vert _{1} \nonumber
\\  & \leq\left\Vert
\mathcal{E}(\varrho_{1})-\rho_{0}\right\Vert _{1}+\left\Vert
\mathcal{E}(\varrho_{2})-\rho_{0}\right\Vert _{1} \nonumber \\
  & \leq2\delta\,,
\end{align}
where on the second line we have applied the definition of
$\delta$-privacy.  It follows that if the scheme is
$\delta$-private, Eve's probability of distinguishing the two
messages is bounded above by $(1+\delta)/2$.

\subsection{Private quantum communication using a private shared Cartesian
frame}

We now determine the superoperator $\mathcal{E}$ that describes
Eve's ignorance of Alice and Bob's private shared Cartesian frame
for states of $N$ spin-1/2 particles.  (Our description applies
equally well to any realization of a qubit that is entirely defined
relative to some reference frame; another example is a single-photon
polarization qubit.)  The transmitted Hilbert space $\mathbb{H}$ in
this case is $(\mathbb{C}^2)^{\otimes N}$. This Hilbert space
carries a tensor power representation $R^{\otimes N}$ of SU(2), by
which an element $\Omega \in$ SU(2) acts identically on each of the
$N$ qubits. For simplicity, we restrict $N$ to be an even integer
for the remainder of this paper, but our main results apply
straightforwardly to all $N$. Then we can decompose
\begin{equation}
  \label{eq:DirectSum}
  (\mathbb{C}^2)^{\otimes N} = \bigoplus_{j=0}^{N/2} \mathbb{H}_j \,,
\end{equation}
where $\mathbb{H}_j$ is the eigenspace of total angular momentum
with eigenvalue $j$.

Each subspace $\mathbb{H}_j$ in the direct sum can be factored into
a tensor product $\mathbb{H}_j = \mathbb{H}_{jR} \otimes
\mathbb{H}_{jP}$, such that SU(2) acts irreducibly on
$\mathbb{H}_{jR}$ and trivially on $\mathbb{H}_{jP}$. Thus,
\begin{equation}
  \label{eq:DirectSumOfProducts}
  (\mathbb{C}^2)^{\otimes N} = \bigoplus_{j=0}^{N/2} \mathbb{H}_{jR}
  \otimes \mathbb{H}_{jP} \, .
\end{equation}
The dimension of $\mathbb{H}_{jR}$ is
\begin{equation}
  \label{eq:MultiplicityR}
  d_{jR}=2j+1\, ,
\end{equation}
and that of $\mathbb{H}_{jP}$ is
\begin{equation}
  \label{eq:Multiplicity}
  d_{jP}=\binom{N}{N/2-j}\frac{2j+1}{N/2+j+1} \, .
\end{equation}

If Alice prepares $N$ qubits in a state $\rho$ and sends them to
Bob, an eavesdropper Eve who is uncorrelated with the private SRF
will describe the state as mixed over all rotations $\Omega \in$
SU(2). Thus, the superoperator $\mathcal{E}$ acting on a general
density operator $\rho$ of $N$ qubits that describes the lack of
knowledge of this private SRF is given by~\cite{BRS03a}
\begin{equation}
  \label{eq:NQubitDecoheringChannel}
  \mathcal{E}(\rho) = \int  R(\Omega)^{\otimes N}
  \rho R^{\dag}(\Omega)^{\otimes N} \,{\rm d}\Omega \, .
\end{equation}
The effect of this superoperator is best seen through use of the
decomposition (\ref{eq:DirectSumOfProducts}) of the Hilbert space.
The action of the superoperator $\mathcal{E}$ can be expressed in
terms of this decomposition as
\begin{equation}
  \label{eq:ActionOfEOnArbitrary}
  \mathcal{E}(\rho) = \sum_{j=0}^{N/2} (\mathcal{D}_{jR} \otimes
  \mathcal{I}_{jP}) (\Pi_j \rho \Pi_j) \, ,
\end{equation}
where $\mathcal{D}_{jR}$ is the completely depolarizing
superoperator on $\mathbb{H}_{jR}$, $\mathcal{I}_{jP}$ is the
identity superoperator on $\mathbb{H}_{jP}$, and $\Pi_j$ is the
projector onto $\mathbb{H}_j$. The subsystems $\mathbb{H}_{jP}$
are called \emph{decoherence-free} or \emph{noiseless}
subsystems~\cite{Kni00} under the action of this superoperator;
states encoded into these subsystems are completely protected from
this decoherence.  In contrast, $\mathcal{E}_N$ is completely
depolarizing on each $\mathbb{H}_{jR}$ subsystem, and thus the
$\mathbb{H}_{jR}$ are called \emph{decoherence-full}
subsystems~\cite{BRS04}.

The largest decoherence-full subsystem occurs for $j_{\rm max}
= N/2$ and has dimension $2j_{\rm max}+1 =N+1$.  As proven
in~\cite{BRS04}, this decoherence-full subsystem defines the
optimally efficient perfectly secure private quantum communication
scheme. Thus, given a private Cartesian frame and the transmission
of $N$ qubits, Alice and Bob can with perfect privacy communicate
$Q(\mathcal{E},0)
= \log_2 (N+1) \sim \log_2 N$ qubits asymptotically.

In contrast, in that same paper it was shown that the private
\emph{classical} capacity using the private shared Cartesian frame
was given by $C(\mathcal{E},0) \sim 3 \log_2 N$. In Appendix
\ref{app:classicalCapacity}, we extend the result to show that
$C(\mathcal{E},\delta) \leq 3(1+\delta)\log_2 N + 3$ for $\delta
\leq 1/2$. The $\delta$-private classical capacity therefore does
not change dramatically when $\delta$ is made non-zero.

\subsection{The working space $\mathbb{H}'$}

To construct a ``working'' Hilbert space on which to investigate
large random subspaces, we use the Hilbert space on which the
states in the private classical communication scheme have support.
This Hilbert space is constructed as follows. Note that for all
$j$ strictly less than the maximum value $N/2$, the
decoherence-free subsystem $\mathbb{H}_{jP}$ is always of
\emph{greater or equal} dimension than the decoherence-full
subsystem $\mathbb{H}_{jR}$. Thus, we will employ irreps up to,
but \emph{not} including, $j=N/2$. Let $j_{\mathrm{min}}<N/2$ be
some fixed irrep.  Our working space $\mathbb{H}^{\prime}$ will
include elements from every irrep in the range
$j_{\mathrm{min}}\leq j<N/2$, that is, for $j\in Y,$ where
\begin{equation}
  Y = \big\{j_{\min},j_{\min}+1,\ldots,N/2-1 \big\}.
\end{equation}
For convenience, we denote the dimension of the decoherence-full
subsystem of the $j_{\mathrm{min}}$ irrep by $D$, that is,
$D\equiv2j_{\mathrm{min}}+1$. Choose a $D-$dimensional subspace
$\mathbb{H}_{jR}^{\prime}$ of $\mathbb{H}_{jR}$ for every $j \in Y$,
and a subspace $\mathbb{H}_{jP}^{\prime}$ of $\mathbb{H}_{jP}$ that
is of dimension $D_\alpha \equiv \left\lfloor \frac{1}{\alpha
}D\right\rfloor ,$ for some parameter $\alpha>1$. Note that such
subspaces always exist because
$\mathrm{dim}\,\mathbb{H}_{jR}=2j+1\geq D$ and
$\mathrm{dim}\,\mathbb{H}_{jP}\geq\mathrm{dim}\,\mathbb{H}_{jR}$ for
all $j\in Y$.

The Hilbert space of interest is then
\begin{equation}
\mathbb{H}^{\prime}=\bigoplus_{j\in
Y}\mathbb{H}_{jR}^{\prime}\otimes \mathbb{H}_{jP}^{\prime}\,,
\end{equation}
with dimensionality $K \equiv \dim\mathbb{H}^{\prime}$ given by
\begin{align}
  K& \sim \frac{1}{\alpha}\sum_{j\in Y}D^{2} \nonumber \\
  & =\frac{1}{\alpha}(N/2-j_{\min})(2j_{\min}+1)^{2} \,.
\end{align}
To maximize this dimension, we choose $j_{\mathrm{min}}$ to be the
integer nearest to $N/3$.  In this case, we have asymptotically
\begin{equation}
  K \sim \frac{2}{27}\frac{1}{\alpha}N^{3}\,.
  \label{dimHprime}
\end{equation}
(More precisely, $K-1$ exceeds the righthand side for sufficiently
large $N$, a result we will use later.)

The superoperator $\mathcal{E}$ maps a state $\varphi$ on
$\mathbb{H}'$ to the state
\begin{equation}
  \label{EonH'}
  \mathcal{E}(\varphi) = \sum_{j\in Y} (I_{\mathbb{H}_{jR}}/d_{jR})
  \otimes {\rm Tr}_{jR}(\Pi_j \varphi \Pi_j) \, ,
\end{equation}
where $I_{\mathbb{H}_{jR}}$ is the identity on $\mathbb{H}_{jR}$.
(Note that the state $\mathcal{E}(\varphi)$ will, in general, have
support outside of $\mathbb{H}'$.) To fully exploit the working
space, we will pursue an encoding such that $\mathcal{E}(\varphi)$
is close to maximally mixed on as large a subspace as possible. To
this end, we define
\begin{equation}\label{Rho0}
    \rho_0 \equiv \sum_{j\in Y} (I_{\mathbb{H}_{jR}}/d_{jR})
  \otimes (I_{\mathbb{H}'_{jP}}/D_{\alpha})\,,
\end{equation}
where $I_{\mathbb{H}'_{jP}}$ is the identity on $\mathbb{H}'_{jP}$.

\section{The main result}

We wish to show that for fixed $\delta$, there exists a private
quantum communication scheme for $\mathcal{E}$ that scales as
$3\log_2 N$. This is achieved by encoding into particular subspaces
of the working space $\mathbb{H}^{\prime}$. Suppose that a subspace
$S \subset \mathbb{H}'$ of the appropriate dimensionality is drawn
at random from some ensemble of subspaces of $\mathbb{H}^{\prime}$.
It is then sufficient to show that the probability that encoding in
$S$ is not $\delta$-private is strictly less than 1, because this
implies that there exist subspaces in the ensemble that do yield
$\delta$-private schemes. Any such subspace can then constitute the
logical Hilbert space $\mathbb{H}_L$ for such a scheme. In this
case, the encoding map $\mathcal{C}$ is simply the embedding map,
which takes states in $S \subset (\mathbb{C}^2)^{\otimes N}$ to
states in $(\mathbb{C}^2)^{\otimes N}$.  Consequently, we leave the
encoding map $\mathcal{C}$ implicit in the rest of the paper.

We shall consider the ensemble of subspaces that is generated by
drawing uniformly at random from among all subspaces of
$\mathbb{H}'$ of a given dimension.  More precisely, we shall take
$S=US_0$ where $S_0$ is a fixed subspace of $\mathbb{H}'$ and $U$
is a unitary on $\mathbb{H}'$ chosen according to the Haar measure
$dU$.  The condition that we require $S$ to satisfy in order to
yield a $\delta$-private scheme is that for all
$|\varphi\rangle\in S$, $\left\Vert
\mathcal{E}(\varphi)-\rho_{0}\right\Vert _{1}\leq\delta$ for
$\rho_0$ given by Eq.~(\ref{Rho0}), so that from Eve's perspective
all the encoded states $\varphi$ are near indistinguishable. This
condition is equivalent to demanding that
\begin{equation}\label{MaxBounded}
  \max_{|\varphi\rangle\in S}\left\Vert \mathcal{E}
  (\varphi)-\rho_{0}\right\Vert _{1}\leq\delta \, ,
\end{equation}
where the maximization is over all pure states in $S$. The
probability that $S$ fails to be $\delta$-private is therefore
\begin{equation}
  \Pr_{S}\left(  \max_{\left\vert \varphi\right\rangle \in S}\left\Vert
  \mathcal{E}(\varphi)-\rho_{0}\right\Vert _{1}>\delta\right)  ,
\end{equation}
where we define the probability $\Pr_{S}(g(S)>\delta)$ that a
randomly-chosen $S$ satisfies some inequality $g(S)>\delta$ by
\begin{equation}
  \Pr_{S}\Big( \,  g(S)>\delta\, \Big)
  \equiv \int_{\{U:g(US_{0})>\delta\}}dU\,.
\end{equation}
For at least one of the $S$ to be $\delta$-private, we require
that
\begin{equation}
  \Pr_{S}\left(  \max_{\left\vert \varphi\right\rangle \in S}\left\Vert
  \mathcal{E}(\varphi)-\rho_{0}\right\Vert _{1}>\delta\right)
  <1 \,,
\end{equation}
for some $\rho_{0}$. The following theorem implies that such
subspaces $S$, with dimension scaling in the desired fashion, do
exist.

\begin{theorem} \label{thm:bigOne}
For the decoherence map $\mathcal{E}$ associated with lacking a
reference frame for $SU(2)$, the condition
\begin{equation}
\Pr_{S}\left(  \max_{\left\vert \varphi\right\rangle \in
S}\left\Vert \mathcal{E}(\varphi)-\rho_{0}\right\Vert
_{1}>\delta\right)  <1 \,,
\end{equation}
holds for sufficiently large $N$, where the probability is with
respect to the unitarily invariant measure on subspaces $S$ of
$\mathbb{H}'$, provided
\begin{equation}
   \log_2\dim S < 3\log_2 N+7/2\log_2 \delta + C'\,
\end{equation}
where $C'$ is a constant.
\end{theorem}

Consider, for example, $1/\delta = {\rm polylog}(N)$, i.e., a
polynomial in $\log(N)$.  Then we can find $S \subset \mathbb{H}'$
with $\Vert \mathcal{E}(\varphi)-\rho_{0}\Vert _{1}\leq\delta$ for
$|\varphi\rangle \in S$ such that
\begin{equation}
  \log_2\dim S \sim 3 \log_2 N,
\end{equation}
recovering the same asymptotic rate as the classical private
capacity. In this case, $Q(\mathcal{E},\delta) \sim
C(\mathcal{E},0)$.

We prove Theorem \ref{thm:bigOne} via a sequence of lemmas. Our
starting point is the following result, known as Levy's
Lemma~\cite{MS86}:

\begin{lemma}
[Levy]\label{lem:Levy} Let $f:S^{k}\rightarrow\mathbb{R}$ be a
continuous real-valued function on the $k$-sphere with Lipschitz
constant $\eta$ with respect to the Euclidean metric. Then, if $x$
is selected at random from $S^{k}$ according to the uniform measure,
\begin{equation}
\Pr_{x}\left(  |f(x)-M|>\gamma\right)  <\exp_2\left(
-C(k-1)\gamma^{2}/\eta
^{2}\right)  ,\label{eqn:Levy}%
\end{equation}
where $C>0$ is a constant and $M$ is a median for $f.$
\end{lemma}

\noindent The function of interest is:
\begin{equation}
  f(\varphi)\equiv\left\Vert
  \mathcal{E}(\varphi)-\rho_{0}\right\Vert_{1}\,.
  \label{defnf}
\end{equation}
Note that the Hilbert space norm on $\mathbb{H}^{\prime}$ is
precisely the Euclidean norm if the Hilbert space is considered as
the real vector space $\mathbb{R}^{2K}$.  The following lemma bounds
the Lipschitz constant of this function.

\begin{lemma}[Lipschitz constant] \label{lem:Lipschitz}
The Lipschitz constant of $f(\varphi)$ is bounded above by $2$.
\end{lemma}

\begin{proof}[Proof] Using the triangle inequality gives
\begin{align}
  |f(\varphi)-f(\tilde{\varphi})|  & = \Bigl\vert \left\Vert
  \mathcal{E} (\varphi)-\rho_0\right\Vert
  _{1}-\left\Vert \mathcal{E}(\tilde{\varphi})
  -\rho_0 \right\Vert _{1}\Bigr\vert \nonumber \\
  & \leq\left\Vert
  \mathcal{E}(\varphi)
  -\mathcal{E}(\tilde{\varphi})\right\Vert_{1}\,.
\end{align}
Because $\mathcal{E}$ is a completely positive trace-preserving
map, and the trace distance is non-increasing under such maps,
\begin{equation}
  \left\Vert
  \mathcal{E}(\varphi)-\mathcal{E}(\tilde{\varphi})\right\Vert
  _{1}\leq\left\Vert
  \varphi-\tilde{\varphi}\right\Vert _{1}\,.
\end{equation}
Combining these inequalities with the fact that
\begin{eqnarray}
 \| |\varphi\rangle - |\tilde{\varphi}\rangle\|_2^2
 &=& 2 - 2 \mbox{Re} \langle\varphi|\tilde{\varphi}\rangle \\
 &\geq& 1 - |\langle\varphi|\tilde{\varphi}\rangle|^2 \\
 &=& \Big( \mbox{$\frac{1}{2}$} \| \varphi - \tilde{\varphi} \|_1 \Big)^2 \,,
\end{eqnarray}
we obtain
\begin{equation}
  |f(\varphi)-f(\tilde{\varphi})|\text{ }\leq2\Vert\left\vert \varphi
  \right\rangle -\left\vert \tilde{\varphi}\right\rangle \Vert_{2} \,,
\end{equation}
which is the desired bound on the Lipschitz constant.
\end{proof}
\medskip

The next corollary, an immediate consequence of Levy's Lemma,
bounds the probability that Eve can distinguish a random state on
$\mathbb{H}'$ from $\rho_0$ substantially better than she can
distinguish states on average.

\begin{corollary}[Concentration of $f$] \label{cor:fConcentration}
Let $|\varphi\rangle$ be chosen at random from the uniform measure
on the unit sphere in $\mathbb{H}'$ and $M$ a median for $f$.  Then
\begin{equation}
 \Pr_\varphi \Big( \big| f(\varphi) - M \big| > \gamma
 \Big) \leq \exp_2\Big( \frac{-C(K-1)\gamma^2}{2}
 \Big).
\end{equation}
\end{corollary}
\begin{proof}[Proof]
Apply Levy to the function $f(\varphi)$ of Eq.~(\ref{defnf}). In
this case, $k=2K-1$ and $\eta \leq 2$.
\end{proof}

Next, we relate the median of $f$ to its mean, which is easier to
estimate.  Write
\begin{equation}
  \mathbb{E}_\varphi f \equiv\int f(\varphi) \,d\nu(\varphi)
\end{equation}
for the expectation of $f$ with respect to the unitarily invariant
measure $d\nu(\varphi)$ on pure states in $\mathbb{H}'$. Let
$A_{\geq}\subset\mathbb{H}^{\prime}$ be the set of points
$\left\vert \varphi\right\rangle $ on the unit sphere for which
$f(\varphi)\geq M$.  By the definition of the median,
\begin{equation}
  \int_{A_{\geq}}f(\varphi)\,d\nu(\varphi)  \geq M\int_{A_{\geq}}d\nu
  (\varphi) =M\cdot\tfrac{1}{2}\,.
\end{equation}
Letting $A_{<}$ be defined analogously, we get
\begin{equation}
  \mathbb{E}_\varphi f
  =\int_{A_{<}}f(\varphi)\,d\nu(\varphi)+\int_{A_{\geq}
  }f(\varphi)d\nu(\varphi) \geq\frac{M}{2},
\end{equation}
because $f(\varphi) \geq 0$.
\begin{lemma}[Expectation of $f$] \label{lem:Expectationf}
The expectation value of $f(\varphi)$ satisfies
\begin{equation}
\mathbb{E}_\varphi f \; \leq\frac{1}{\sqrt{\alpha}}.
\end{equation}
\end{lemma}

The proof is supplied in Appendix \ref{app:calcExpectation}, but
can be understood intuitively in terms of the action of
$\mathcal{E}$ on $\mathbb{H}^{\prime}$. If the subspaces
$\mathbb{H} _{jP}^{\prime}$ were $1$-dimensional, then by virtue
of the fact that the $\mathbb{H}_{jR}$ are decoherence-full, we
would have complete decoherence on
$\mathbb{H}_{jR}\otimes\mathbb{H}_{jP}^{\prime}$. Because
$1/\alpha\sim\dim\mathbb{H}_{jP}^{\prime}/\dim\mathbb{H}_{jR}^{\prime}$,
the larger the value of $\alpha,$ the smaller the dimension of
$\mathbb{H}_{jP}^{\prime}$ relative to $\mathbb{H}_{jR}^{\prime}$,
and the less distinguishable on average are states on
$\mathbb{H}_{jR}\otimes\mathbb{H}_{jP}^{\prime}$ subsequent to the
action of $\mathcal{E}$. One might expect that states could be
distinguished by their relative supports on the different irreps
$j \in Y$, because these supports are invariant under the action
of $\mathcal{E}$. However, the proof demonstrates that because we
use only sufficiently large irreps, all encoded states will have
similar supports on all irreps, and thus not be significantly more
distinguishable than if a single irrep had been used.

We note that the proof of the lemma requires that, within each
irrep $j\in Y$, the dimension $D_\alpha$ of $\mathbb{H}'_{jP}$ be
much smaller than the dimension $D$ of the decoherence-full
subsystem $\mathbb{H}_{jR}$. For this reason, our result does not
apply directly to cryptography using a U(1) phase reference, for
which the decoherence-full subsystems are all one-dimensional.
However, for any other reference frame that satisfies this
condition, our results should be directly applicable.

We conclude that the median $M$ is upper bounded by
$2/\sqrt{\alpha}$ which, using Corollary~\ref{cor:fConcentration},
leads to the result
\begin{multline}
  \label{OldTheoremI4}
  \Pr_{\varphi}\left(  \left\Vert
  \mathcal{E}(\varphi)-\rho_{0}\right\Vert _{1}>\gamma +
  \frac{2}{\sqrt{\alpha}} \right) \\
  \leq\exp_2\left( -\frac{C}{2}(K-1)\gamma^{2}\right)\,.
\end{multline}

This inequality is sufficiently strong that we will be able to use
it to conclude that large subspaces of $\mathbb{H}'$ have the
property that the distinguishability of \emph{all} states in the
subspace are bounded.

\begin{lemma}[Existence of good subspaces]  Let $S_0 \subset
\mathbb{H}'$ be a fixed subspace and $|\varphi_0\rangle$ a fixed
state on $S_0$. Let $S = U S_0$ be a random subspace obtained from
$S_0$ using a Haar-distributed unitary $U$ on $\mathbb{H}'$. Then,
for any $\delta > 0$ and $0 <  \varepsilon < 1/2$,
\begin{multline}
  \Pr_{S}\Bigl(  \max_{\left\vert \varphi\right\rangle \in
  S} f(\varphi) >\delta\Bigr) \\
  \leq\left(  \frac{5}{\varepsilon}\right) ^{2\dim
  S}\Pr_{U}\Bigl( f(U |\varphi_0\rangle) >\delta
  -\varepsilon\Bigr).
\end{multline}
\end{lemma}

\begin{proof}[Proof]
Fix an $\epsilon/2$-net $\mathcal{N}_{0}$ for the unit sphere of a
fixed subspace $S_{0}$ of $\mathbb{H}^{\prime}$ with the Hilbert
space norm. The net can be chosen such that the number of elements
in the net satisfies $|\mathcal{N}_{0}| \leq(5/\epsilon)^{2\dim
S_{0}}$. (See~\cite{HLW04} a proof of this fact.) By definition,
given any $\left\vert \varphi\right\rangle \in S_{0}$, there exists
a state $\left\vert \tilde{\varphi}\right\rangle \in\mathcal{N}_{0}$
such that $\Vert\left\vert \varphi\right\rangle -\left\vert
\tilde{\varphi}\right\rangle \Vert_{2} \leq\epsilon/2$.  By
Lemma~\ref{lem:Lipschitz}, this implies that $|f(\varphi
)-f(\tilde{\varphi})|\leq\epsilon$.

Now choose a random subspace $S=US_{0}$ using a Haar-distributed
unitary.  This unitary $U$ maps the net $\mathcal{N}_{0}$ for
$S_{0}$ into a net $\mathcal{N}$ for $S$.  Let $\left\vert
\varphi^{\ast}\right\rangle $ be defined by
\begin{equation}\label{fstar}
  f(\varphi^{\ast })= \underset{\left\vert \varphi\right\rangle \in
  S}{\max}\,f(\varphi) \,.
\end{equation}
By definition, there exists a state $\left\vert
\tilde{\varphi}^{\ast }\right\rangle \in\mathcal{N}$ such that
$\Vert\left\vert \varphi^{\ast }\right\rangle -\left\vert
\tilde{\varphi}^{\ast}\right\rangle \Vert_{2} \leq\epsilon/2$, and
consequently $|f(\varphi^{\ast})-f(\tilde{\varphi}^{\ast
})|\leq\epsilon$.  It follows that if $f(\varphi^{\ast})>\delta$,
then $f(\tilde{\varphi}^{\ast})>\delta-\epsilon$.  Therefore, if
\begin{equation}
    \underset{\left\vert \varphi\right\rangle \in
  S}{\max}\,f(\varphi) >\delta\,, \quad\mbox{then}\quad
%\end{equation}
%then
%\begin{equation}
    \underset{\left\vert \tilde{\varphi}\right\rangle
    \in\mathcal{N}}{\max}\,f(\tilde{\varphi} )>\delta-\epsilon \,.
\end{equation}
Finally, if $x$ implies $y,$ then $\Pr(x)\leq\Pr(y),$ so we conclude
that
\begin{equation}
  \Pr_{S}\left(  \underset{\left\vert \varphi\right\rangle \in S}{\max
  }\,f(\varphi)>\delta\right)
  \leq\Pr _{U}\left(  \underset{\left\vert
  \tilde{\varphi}\right\rangle \in
  \mathcal{N}_{0}}{\max}\,f(U\left\vert \tilde{\varphi}\right\rangle
  )>\delta-\epsilon\right)
\end{equation}
where $\Pr_{U}$ reminds the reader that we are varying over
unitaries.  We then have
\begin{align}
  \Pr_{U}\Bigl(  \underset{\left\vert \tilde{\varphi}\right\rangle
  \in\mathcal{N}_{0}}{\max}\, & f(U\left\vert
  \tilde{\varphi}\right\rangle )>\delta-\epsilon\Bigr) \nonumber\\
  & \leq\sum_{\left\vert \tilde{\varphi }\right\rangle
  \in\mathcal{N}_{0}}\Pr_{U}\left(  \,f(U\left\vert
  \tilde{\varphi}\right\rangle )>\delta-\epsilon\right) \nonumber\\
  & =\left\vert \mathcal{N}_{0}\right\vert \Pr_{U}\left(
  \,f(U\left\vert \tilde{\varphi}_{0}\right\rangle
  )>\delta-\epsilon\right)\,,
\end{align}
where the first inequality is the union bound for probabilities and
the second line follows from the fact that the expression inside the
sum over $\left\vert \tilde{\varphi}\right\rangle $ is independent
of $\left\vert \tilde{\varphi }\right\rangle $ ($\left\vert
\tilde{\varphi}_{0}\right\rangle $ is an arbitrary state in
$\mathbb{H}^{\prime}$). $\ $Recalling that $|\mathcal{N} _{0}|$
$\leq(5/\epsilon)^{2\dim S_{0}}$ establishes what we set out to
prove. \end{proof} \medskip

Using the lemma together with Eq.~(\ref{OldTheoremI4}), we obtain
\begin{multline}
  \Pr_{S}\left(  \max_{\left\vert \varphi\right\rangle \in
  S}\left\Vert \mathcal{E}(\varphi)-\rho_{0}\right\Vert
  _{1}>\delta\right) \\
  \leq\left(  \frac{5}{\varepsilon}\right) ^{2\dim
  S}\exp_2\left( -\frac{C}
  {2}(K-1)(\delta-\varepsilon-\frac{2}{\sqrt{\alpha}}
  )^{2}\right)\,.
\end{multline}
If $\dim S$ is chosen such that the right hand side is bounded
away from $1,$ then the left hand side will also be bounded away
from $1,$ and there will exist a $\delta$-private encoding into a
subspace $S.$ We will therefore seek the largest value of $\dim S$
that satisfies the inequality
\begin{equation}
  \left(  \frac{5}{\varepsilon}\right)  ^{2\dim S}<\exp_2\left(
  \frac{C}{2}
  (K-1)(\delta-\varepsilon-\frac{2}{\sqrt{\alpha}}
  )^{2}\right),
\end{equation}
or equivalently
\begin{equation}
    \label{dimSbound1}
  \dim S  <\frac{\ln 2}{\ln\left(  \frac{5}{\varepsilon}\right)
  }\frac{C}{4}
  (K-1)(\delta-\varepsilon-\frac{2}{\sqrt{\alpha}})^{2}.
\end{equation}
Given that $\ln x\leq \sqrt{x}$, we have $1/\ln\left(
\frac{5}{\varepsilon}\right) \geq1/\sqrt{\frac{5}{\varepsilon}}$
and any $S$ satisfying
\begin{equation}
    \label{dimSbound2}
  \dim S<\sqrt{\frac{\varepsilon}{5}}\frac{C\ln
  2}{4}(K-1)(\delta-\varepsilon-\frac{2}{\sqrt{\alpha}})^{2}\,,
\end{equation}
will also satisfy Eq.~(\ref{dimSbound1}). Using the expression for
$K$ in Eq.~(\ref{dimHprime}), it is sufficient to require that
\begin{equation}
  \dim
  S<\frac{C\ln 2}{54\sqrt{5}}\frac{1}{\alpha}N^{3}(\delta-\varepsilon-\frac
  {2}{\sqrt{\alpha}})^{2}\sqrt{\varepsilon}\,
\end{equation}
for sufficiently large $N$. If we choose $\varepsilon
= \delta/3$ and $\alpha = 36/\delta^2$, then this expression
reduces to
\begin{equation}
  \dim S<\frac{C\ln 2}{5832\sqrt{15}}N^3 \delta^{7/2}\,.
\end{equation}
It is therefore possible to choose $S$ such that $f(\varphi) \leq
\delta$ for all $|\varphi\rangle \in S$ whenever
\begin{equation}
  \log_2\dim S < 3\log_2 N+7/2\log_2 \delta + C'\,
\end{equation}
where $C' = \log_2[(C\ln2)/(5832\sqrt{15})]$, completing the proof
of Theorem \ref{thm:bigOne}.

\section{Discussion}

We have seen that for fixed $\delta > 0$, the $\delta$-private
quantum capacity of a secret SU(2) reference frame is at least
three times as large as its perfectly private quantum capacity.
Indeed, the relaxation of the security requirement to $\delta
> 0$ causes the private quantum capacity to jump almost to the
value of the perfectly private classical capacity, which is
approximately $3 \log_2 N$, and within a factor of $1+\delta$ of the
$\delta$-private classical capacity. In earlier work, a similar
relaxation of the security condition in the quantum one time pad led
to a doubling of the private quantum capacity of a shared secret key
string~\cite{HLSW04} as well as a similar doubling of the capacity
of a maximally entangled state~\cite{HHL04,HLW04}. The tripling of
the capacity seen here, however, is unusual and reflects the
particular structure of the tensor power representation of SU(2).

Because the private capacity of a shared reference frame is
proportional to $\log_2 N$ rather than $N$, however, the values of
$\delta$ which provide an improvement over the perfectly private
schemes is quite restricted. From Theorem \ref{thm:bigOne}, we see
that for sufficiently large $N$,
\begin{equation}
Q(\mathcal{E},\delta) \geq 3 \log_2 N + \frac{7}{2} \log_2 \delta
+ C'
\end{equation}
for some constant $C'$. In order to improve upon the perfectly
private scheme, we require that $Q(\mathcal{E},\delta) > \log_2
N$, which implies that $1/\delta \in O(N^{4/7})$. In particular,
our construction does not allow $\delta$ to be an exponentially
decreasing function of $N$, which would obviously be more
desirable for cryptographic applications.

Some questions remain about the optimality of the private quantum
communication schemes we have presented here. In particular, our
upper bounds on the private quantum capacity do not exclude the
possibility that $\delta$ could be made to shrink exponentially
with $N$ while maintaining a number of qubits sent scaling as $3
\log_2 N$. Also, we have not attempted to construct
$\delta$-private classical communication schemes meeting the upper
bound of Theorem \ref{thm:privateCapacity} in
Appendix~\ref{app:classicalCapacity}.

Finally, we note that a shared Cartesian frame is not the only
possible form of a shared reference~\cite{BRS04}, and it is useful
to consider other practical examples such as a shared phase
reference, shared direction, or reference ordering.  These examples
have different Hilbert-space structures arising from their group
representation theory, and in general will result in different
relations between their private classical and quantum capacities. We
note that our technique should apply directly to cryptography using
a reference frame for $U(K)$, with $K \geq 2$, because the Hilbert
space structures for these groups satisfy the conditions required
for our proof.  Whether similar differences between
perfectly-private and $\delta$-private capacities can be found for
other reference frames is an open question.

\begin{acknowledgments}
The authors gratefully acknowledge J. Emerson, D. Gottesman, and T.
Rudolph for helpful discussions. SDB acknowledges support from the
Australian Research Council, PH appreciates the support of the
Canadian Institute for Advanced Research, and both PH and RWS are
grateful for support from the Natural Sciences and Engineering
Research Council of Canada.
\end{acknowledgments}

\appendix

\section{$\delta$-private classical capacity}
\label{app:classicalCapacity}

\begin{theorem} \label{thm:privateCapacity}
For $\delta \leq 1/2$, the $\delta$-private
classical capacity satisfies $C(\mathcal{E},\delta) \leq
3(1+\delta)\log_2 N + 3$.
\end{theorem}
\begin{proof}
Suppose we have a $\delta$-private classical communication scheme
for $\mathcal{E}$ consisting of $m$ states on $\mathbb{H}$. If
such a scheme exists, then there is also an $m$-state scheme using
pure states, which we will label $\{ |\psi_i\rangle \}_{i=1}^m$.
We will use the privacy condition to find a small subspace of
$\mathbb{H}$ such that these states are almost entirely contained
within the subspace. Combining the Holevo bound with the fact that
the original states were all distinguishable will then lead to an
upper bound on $m$, the number of states in the scheme.

Let $\mathbb{H}_{jQ}$ be the subspace of $\mathbb{H}_{jP}$
corresponding to the non-zero eigenvalues of $\mathrm{Tr}_{jR} [
\Pi_j \psi_1 \Pi_j ]$ and let $\Pi_{jQ}$ be the projector onto
$\mathbb{H}_{jQ}$. It follows from the Schmidt decomposition for
$\Pi_j |\psi_1\rangle$ that $\dim \mathbb{H}_{jQ} \leq \min(
d_{jR}, d_{jP})$. Also let $\Pi'
= \sum_j \Pi_{jR} \otimes \Pi_{jQ}$, where $\Pi_{jR}$ is the
projector onto $\mathbb{H}_{jR}$. Observe that for any $\psi_i$,
\begin{eqnarray}
\mathrm{Tr}[ \Pi' \psi_i \Pi' ]
 = \mathrm{Tr}[ \mathcal{E}(\Pi' \psi_i \Pi') ]
 = \mathrm{Tr}[ \Pi' \mathcal{E}(\psi_i) \Pi' ]
 \label{eqn:handyCommutation}
\end{eqnarray}
because $\mathcal{E}$ is trace-preserving and because projection
by $\Pi'$ commutes with $\mathcal{E}$. By the privacy condition,
however,
\begin{eqnarray}
\delta
 &\geq& \| \mathcal{E}(\psi_1) - \mathcal{E}(\psi_i) \|_1 \\
 &\geq& 2 \{ \mathrm{Tr}[\Pi' \mathcal{E}(\psi_1) \Pi'] -
    \mathrm{Tr}[\Pi' \mathcal{E}(\psi_i) \Pi']\} \\
 &=& 2 \{ 1 - \mathrm{Tr}[\Pi' \mathcal{E}(\psi_i) \Pi'] \}.
 \label{eqn:truncatedPrivacy}
\end{eqnarray}
The second inequality holds because $\|X\|_1 = 2 \max_P
\mathrm{Tr}[PX]$ for traceless, Hermitian $X$, where the
optimization is over projectors of all ranks. (See, for example,
\cite{Nie00}.) Combining (\ref{eqn:truncatedPrivacy}) with
(\ref{eqn:handyCommutation}) shows that $\mathrm{Tr}[ \Pi' \psi_i
\Pi' ] \geq 1 - \delta/2$. Thus the states $\{ |\psi_i\rangle
\}_{i=1}^m$ are essentially contained within the subspace defined
by $\Pi'$.

Now consider the set of states $\{|\psi_i'\rangle\}_{i=1}^m$,
where
\begin{equation}
|\psi_i'\rangle
 = \frac{\Pi'|\psi_i\rangle}{\sqrt{\mathrm{Tr}[\Pi'\psi_i\Pi']}}.
\end{equation}
Because $|\langle\psi_i|\psi_i'\rangle|^2 =
\mathrm{Tr}[\Pi'\psi_i\Pi']$, performing the measurement
$\{|\psi_j\rangle\langle\psi_j|\}_{j=1}^m$ on the set of states
$\{|\psi_i'\rangle\}_{i=1}^m$ will correctly identify the state with
probability at least $1-\delta/2$. Assume a state $|\psi_i'\rangle$
is chosen from the uniform distribution. By Fano's
inequality~\cite{CT91},
\begin{equation}
H(i|j) \leq 1 + \frac{\delta}{2} \log_2 m,
\end{equation}
where $H$ is the Shannon conditional entropy function, which in
turn implies that
\begin{equation}
I(i;j) \geq (1-\delta/2)\log_2 m - 1,
\end{equation}
where $I$ is the mutual information function. Because all the
states $|\psi_i'\rangle$ are contained in the support of $\Pi'$,
however, the Holevo bound~\cite{H73} implies that $I(i;j)$ is no
more than the logarithm of $\mathrm{rank} \, \Pi'$, which
satisfies
\begin{equation}
\mathrm{rank} \, \Pi'
 \leq \sum_j d_{jR} \cdot \min( d_{jR}, d_{jP}).
\end{equation}

In the case of a private shared SU(2) reference frame, for which
$d_{jR} = 2j+1$,
\begin{equation}
\mathrm{rank} \, \Pi' \leq (N/2+1)(N+1)^2 \leq 2N^3,
\end{equation}
where the second inequality holds for all $N \geq 2$. This implies
that
\begin{eqnarray}
\log_2 m
 &\leq& \frac{3 \log_2 N + 2}{1 - \delta/2} \\
 &\leq& 3(1+\delta) \log_2 N + 3,
\end{eqnarray}
provided $\delta \leq 1/2$.
\end{proof}

\section{Proof of Lemma~\ref{lem:Expectationf}}
\label{app:calcExpectation}

The map $\mathcal{E}$ depolarizes each of the systems
$\mathbb{H}_{jR}$ but for the purposes of calculation, it is
easier to simply discard them. In the proof, therefore, we will
work with the space $\mathbb{H}_{P}^{\prime}=\oplus_{j\in
Y}\mathbb{H}_{jP}^{\prime}$, which has dimension
$d_{P}\equiv\dim\mathbb{H}_{P}^{\prime}$. Observe that if we
introduce
\begin{equation}
  \mathcal{F}(\rho)=\sum_{j \in
  Y}\mathrm{Tr}_{jR}(\Pi_{j}\rho\Pi_{j})\,,
\end{equation}
which gives a normalized state on $\mathbb{H}_P'$, then
\begin{equation}
  \Vert\mathcal{E}(\varphi)-\rho_{0}\,\Vert_{1}=\Vert\mathcal{F}(\varphi
  )-\varrho_0\Vert_{1}\,,
\end{equation}
where $\varrho_0 = I_P/d_P$ is the normalized identity operator on
$\mathbb{H}_{P}^{\prime}$.

Using $\Vert X\Vert_{1}\leq\sqrt{\mathrm{rank}X}\Vert X\Vert_{2}$
gives
\begin{equation}
  \Vert\mathcal{F}(\varphi)-\varrho_0 \,\Vert_{1}\leq\sqrt{d_{P}}
  \Vert\mathcal{F}(\varphi)-\varrho_0 \,\Vert_{2}\,.
\end{equation}
We therefore have
\begin{align}
  \mathbb{E}_\varphi f  & \leq\sqrt{d_{P}}\int
  \Vert\mathcal{F}(\varphi)-\varrho_0 \,\Vert_{2}\,d\nu(\varphi) \nonumber \\
  & =\sqrt{d_{P}}\int \sqrt{\mathrm{Tr}\left[
  \mathcal{F}
  (\varphi)^{2}-
 \mathcal{F}(\varphi)/d_{P}+I_{P}/d_{P}^{2}\right]}\,d\nu(\varphi)\,.
\end{align}
Using the normalization $\mathrm{Tr}\mathcal{F}(\varphi)=1$ and the
concavity of the square root function, this expression reduces to
\begin{equation}
\mathbb{E}_\varphi f \leq \sqrt{ \int d_P
\mathrm{Tr}[\mathcal{F}(\varphi)^{2}]\,d\nu(\varphi)-1}.
\end{equation}
It therefore suffices to evaluate
\begin{multline}
  \int \mathrm{Tr}[\mathcal{F}(\varphi)^{2}] \,d\nu(\varphi)\\
  =\int \mathrm{Tr}\Bigl[ \Bigl(  \sum_{j\in Y}\mathrm{Tr}_{jR}\left(
  \Pi_{j} \varphi\Pi_{j}\right)  \Bigr)^{2} \Bigr] \,d\nu
  (\varphi)\,. \label{eqn:integralSimple}
\end{multline}
Because $\Pi_j$ has the form $\Pi_j = \Pi_{jR} \otimes \Pi_{jP}$,
where $\Pi_{jR}$ and $\Pi_{jP}$ are the projectors onto
$\mathbb{H}_{jR}$ and $\mathbb{H}_{jP}$ respectively,
$\mathrm{Tr}_{jR} ( \Pi_j \varphi \Pi_j )$ and $\mathrm{Tr}_{kR}
(\Pi_k \varphi \Pi_k)$ have orthogonal supports, implying that
\begin{equation} \label{eqn:orthogProjectors}
  \mathrm{Tr}\Bigl[ \Bigl(  \sum_{j \in Y}\mathrm{Tr}_{jR}\left(
  \Pi_{j} \varphi\Pi_{j}\right)  \Bigr)^{2} \Bigr] \\
  = \mathrm{Tr}\Bigl[ \sum_{j \in Y} \Bigl(\mathrm{Tr}_{jR}\left(
  \Pi_{j} \varphi\Pi_{j}\right) \Bigr)^2 \Bigr] \,.
\end{equation}
To evaluate the resulting integral, fix bases
$\{|m\rangle\}_{m=1}^D$ and $\{|l\rangle\}_{l=1}^{D_\alpha}$ for
the spaces $\mathbb{H}_{jR}'$ and $\mathbb{H}_{jP}'$ respectively.
(Note that we identify bases labelled by different values of $j$.)
Also let $|\varphi_0\rangle = |j_0 m_0 l_0\rangle$ for some fixed
values of $j_0$, $m_0$ and $l_0$. Using
Eq.~(\ref{eqn:orthogProjectors}) and making use of the invariance
of the measure, we can expand the integral of
Eq.~(\ref{eqn:integralSimple}) as
\begin{widetext}
\begin{equation}
  \int_{{\rm U}(K)} \mathrm{Tr}[\mathcal{F}(U\varphi_{0}U^\dagger)^{2}]\,dU =
\sum_{j \in Y}
  \sum_{m,m^{\prime}=1}^D \sum_{l,l^{\prime}=1}^{D_\alpha} \int_{{\rm
U}(K)} U_{jml,j_0 m_{0}l_{0}}
  U_{jml^{\prime},j_0 m_{0}l_{0}}^{\ast}U_{jm^{\prime}l^{\prime
  },j_0 m_{0}l_{0}}U_{jm^{\prime}l,j_0 m_{0}l_{0}}^{\ast}\,dU\,,
\end{equation}
which can be evaluated using the identity (see, for example,
\cite{AL03})
\begin{equation}
  \int_{{\rm U}(K)}U_{ij}U_{kl}^{\ast}U_{mn}U_{pq}^{\ast}\,dU=\frac{1}{K^{2}
  -1}\left\{
  \delta_{ij,kl}\delta_{mn,pq}+\delta_{ij,pq}\delta_{kl,mn}-\frac
  {1}{K}\delta_{ik}\delta_{jq}\delta_{ln}\delta_{mp}-\frac{1}{K}\delta
  _{ip}\delta_{jl}\delta_{km}\delta_{nq}\right\}
  \,.\label{eqn:fourUnitaries}
\end{equation}
We obtain
\begin{align}
  \int_{{\rm U}(K)}  \mathrm{Tr}[\mathcal{F}(U\varphi_{0}U^\dagger)^{2}]\,dU  &
=\sum_{j \in Y}
  \sum_{m,m^{\prime}=1}^D \sum_{l,l^{\prime}=1}^{D_\alpha}
\frac{1}{K(K+1)}\left\{
  \delta_{l,l^{\prime}}+\delta_{m,m^{\prime}}\right\}
  \\
  & =\frac{\sum_{j\in Y}(D^{2}D_\alpha + D_\alpha^{2}D)}{K(K+1)}\,.
\end{align}
\end{widetext}
Substituting this back into the expression for $\mathbb{E}_\varphi
f$ yields
\begin{equation}
  \mathbb{E}_\varphi f \leq
  \sqrt{\frac{d_{P}}{K(K+1)} \Bigl( \sum_{j\in Y}(D^{2}D_\alpha + D_\alpha^{2}D)
\Bigr) -1}
  \,.
\end{equation}
Recalling that $d_P = \sum_{j\in Y}D_\alpha$ and $K = \sum_{j\in
Y}D_\alpha D$, we get $\mathbb{E}_\varphi
f\leq\sqrt{{D_\alpha}/{D}}$. Because $D_\alpha \leq
\frac{1}{\alpha} D$, we have the desired inequality:
\begin{equation}
\mathbb{E}_\varphi f \leq \sqrt{\frac{1}{\alpha}}\,.
\end{equation}

\end{document}